\documentclass[prl,twocolumn,showpacs]{revtex4}
\usepackage{amsmath}
\usepackage[dvips,xdvi]{graphicx}
\usepackage{amssymb}
\usepackage{epsfig}
\usepackage{color}

\begin{document}
\title{Number Fluctuations and Energy Dissipation in Sodium Spinor Condensates}
\author{Y. Liu$^1$}
\email{yingmei.liu@nist.gov}
\author{E. Gomez$^2$}
\author{S. E. Maxwell$^1$}
\author{L. D. Turner$^3$}
\author{E. Tiesinga$^1$}
\author{P. D. Lett$^1$}
\email{paul.lett@nist.gov}
\affiliation{$^1$Joint Quantum
Institute, University of Maryland and \\National Institute of
Standards and Technology, Gaithersburg, MD 20899,
USA\\$^2$Instituto de F{\'{i}}sica, Universidad Aut{\'{o}}noma de
San Luis Potos{\'{i}}, San Luis Potos{\'{i}} 78290,
Mexico\\$^3$School of Physics, Monash University, Victoria 3800,
Australia}
\date{\today}

\begin{abstract}
We characterize fluctuations in atom number and spin populations
in $F=1$ sodium spinor condensates. We find that the fluctuations
enable a quantitative measure of energy dissipation in the
condensate. The time evolution of the population fluctuations
shows a maximum. We interpret this as evidence of a
dissipation-driven separatrix crossing in phase space. For a given
initial state, the critical time to the separatrix crossing is
found to depend exponentially on the magnetic field and linearly
on condensate density. This crossing is confirmed by tracking the
energy of the spinor condensate as well as by Faraday rotation
spectroscopy. We also introduce a phenomenological model that
describes the observed dissipation with a single coefficient.

\end{abstract}
\maketitle

The transition from a thermal atomic gas to a Bose-Einstein
condensate (BEC) is marked by the appearance of a scalar order
parameter. Spinor BECs have an additional spin degree of freedom
which results in a vector order parameter. The increase in
complexity leads to the formation of spin
domains~\cite{chapman1,zhang05,ketterle1}, the appearance of novel
phases~\cite{barnett06} and the possibility of high spatial
resolution magnetometry~\cite{vengalattore07}. Spin-1 spinor BECs
have been studied with $^{23}$Na atoms that show antiferromagnetic
interactions~\cite{ketterle1,ketterle2,lett,faraday} and with
$^{87}$Rb atoms that show ferromagnetic
interactions~\cite{chapman1,chapmanQ,lyou1,f2Sengstock2,bloch,gerbier06}.
A remarkable result is the observation of spin population
oscillations that appear when the system is taken out of
equilibrium in the presence of a magnetic
field~\cite{chapman1,lett}. An interplay between the quadratic
Zeeman energy and a spin-dependent interaction energy determines
the oscillation frequency. The oscillations are nearly harmonic
except near a separatrix in phase space where the period
diverges~\cite{lyou1,lett,faraday}. The system can be forced onto
either the low or high energy side of the separatrix using the
magnetic field strength~\cite{lett,lyou1}, the BEC
density~\cite{chapman1}, or the balance among spin
states~\cite{faraday}.

Quantum optical effects in spinor BECs are now being actively
studied. Recent experiments observed the spin-mixing analogue of
parametric amplification~\cite{ErtmerArxiv2009}. Number
fluctuations and spin oscillations were investigated in a
ferromagnetic Rb BEC in Refs.~\onlinecite{chapmanQ}
and~\onlinecite{youQ}. The observed spin oscillations damped and
the system reached a steady state, while the fluctuations
saturated. Non-dissipative theories of quantum effects in number
fluctuations~\cite{youQ,bigelowDis1,bigelowDis2,Ho} show that such
damped spin oscillations can be produced by dephasing from quantum
fluctuations or from number and phase fluctuations in the initial
state.

We report observations of the dynamics of atom number fluctuations
in an antiferromagnetic $^{23}$Na spinor BEC which show strong
evidence of energy dissipation. Atom number fluctuations of the
spin projections are extracted from a series of measurements on
condensates which spin-mix while slowly evolving to the ground
state. Because of the dissipation, a spinor BEC with given initial
conditions and sufficiently high energy will cross the separatrix
at a critical time, $t_c$. The time evolution of the population
fluctuations unambiguously identifies $t_c$. We have developed a
dissipation model that includes a single phenomenological
coefficient, is classical, and does not require the intrinsic
quantum fluctuations that have been used to describe ferromagnetic
Rb spinor systems. Other dissipation mechanisms have been
suggested in~\cite{bloch,SengstockDis1,SengstockDis2}, but do not
explain our data. Mean-field simulations at zero magnetic field
including finite temperature effects in
Ref.~\onlinecite{MorenoCardoner} indicate that thermal excitations
play a prominent role in the dissipation.

We show that a non-dissipative quantum model does not account for
our data. This model predicts a damped spin oscillation due to
quantum dephasing similar to the ferromagnetic case. Its predicted
steady states, however, are very different from the experimental
observations in this paper and Ref.~\onlinecite{faraday}. It is
interesting to note that a spinor BEC represents a nominally
isolated quantum system that shows dissipation and does not
display quantum rephasing over our observation times.

The setup is similar to that of our previous work \cite{faraday}.
We create a $F=1$ spinor BEC of $N=1.50(3)\times10^5$ $^{23}$Na
atoms through 6~s forced evaporation in a crossed optical dipole
trap using a multi-mode fiber laser at 1070~nm (all quoted
uncertainties are one standard deviation, combined statistical and
systematic). We apply a weak magnetic field gradient during the
evaporation to fully polarize the atoms to the $| F=1, m_F = +1
\rangle$ state. The final trap oscillation frequencies are
$\omega_{0}(\sqrt{2},1,1)$ in the three spatial directions. We
have performed experiments with measured values of
$\omega_0/(2\pi)$~=~154(5)~Hz, 220(7)~Hz, and 305(9)~Hz. This
corresponds to mean Thomas-Fermi radii of 6$~\mu$m to 8$~\mu$m.
The density of the BEC is changed only by changing the trap
frequency and not by reducing the atom number, as this causes a
large loss in signal.

To prepare the initial state, we turn off the magnetic field
gradient and ramp to a magnetic field, $B$, less than 61~$\mu$T.
We apply an rf pulse resonant with the linear Zeeman splitting
(frequencies of hundreds of kHz) to rotate the atomic spin. All of
our experiments start with the same initial atomic state which has
$\rho_{+1}=\rho_{-1}=1/4$ and $\rho_0=1/2$, where the $\rho_{m_F}$
are fractional populations for the three Zeeman sublevels
($m_F=0,\pm 1$).  This state has zero magnetization, where the
magnetization is defined as $m=\rho_{+1} - \rho_{-1}$. We use two
methods to detect spin mixing dynamics: Faraday rotation
spectroscopy and Stern-Gerlach separation combined with absorption
imaging (SG-AI). Faraday rotation spectroscopy can be used for
continuous observation of spin oscillations of a single BEC over
short time scales, while SG-AI can directly measure spin
populations, albeit destructively~\cite{faraday}.

The Mandel Q parameter is a common way to quantify fluctuations in
quantum systems, with $Q > 0~(< 0 )$ representing super- (sub-)
Poissonian distributions~\cite{mandel}. We use a modified Q
parameter to characterize population fluctuations during spin
mixing. This Q parameter of $\rho_0$ is defined as~\cite{youQ}
\begin{equation}
Q=\langle N\rangle
\frac{\langle\triangle\rho_0^2\rangle}{\langle\rho_0\rangle}-1,
\label{mandelQ}
\end{equation}
where $\langle N\rangle$ is the mean value of the atom number in
the BEC. $\langle\triangle\rho_0^2\rangle$ and
$\langle\rho_0\rangle$ correspond to the variance and the mean
value of $\rho_0$, respectively. At each delay time after
initialization, we extract the variance from 25-30 repeated SG-AI
measurements. We use $\rho_0$ rather than the population, because
measurements of $\rho_0$ and $m$ are less sensitive to the 2\%
fluctuations in the initial $N$.

\begin{figure}[t]
\includegraphics[width=85mm]{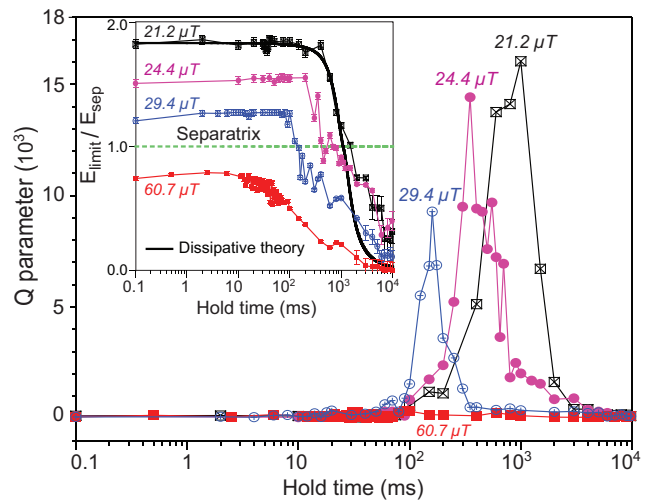}
\caption{(Color online) Time evolution of $Q$ with $\langle
n\rangle=1.37(6)\times10^{14}$ cm$^{-3}$ ($c/h=$ 33(1) Hz) at four
magnetic fields as indicated in the figure. Inset: Time evolution
of $E_{\rm limit}$/$E_{\rm sep}$ for the same data (see text).
Good agreement between the predictions of Eqs.~(3) and the data is
found using a $\beta$ that falls on the line shown in the inset of
Fig.~\ref{beta}.  An example curve is shown for $B=21.2~\mu$T
(thick black line).} \label{qfactor}
\end{figure}

Theoretically, our initial state, prepared from a single component
BEC, should be a coherent state with a Poissonian atom number
distribution ($Q=0$). The observed $Q$ at $t=0$ is equal to 4,
while the minimum observable $Q$ at other times depends on the
populations, due to technical noise in atom counting. In
particular, when one of the spin populations is close to zero the
minimum value of, and the error in, $Q$ are larger.

Figure~\ref{qfactor} shows the time evolution of $Q$ for $m=0$ at
four magnetic field strengths. $Q$ has a value equal to the
experimental limit at $t=0$, increases to a peak at $t_c$, then
decreases back to the experimental limit within 10 s. We find that
at $t=10$~s all the remaining atoms are in the $m_F=0$ state
($\approx$20\% are lost). A Gaussian fit to $Q(t)$ is applied to
extract $t_c$. We observe that the value of $t_c$ decreases with
increasing field. For high fields (e.g., $B=60.7$~$\mu$T in
Fig.~\ref{qfactor}) where the system is on the low energy side of
the separatrix at $t=0$, $Q(t)$ stays at the experimental limit
during the whole evolution.

The single mode approximation (SMA)~\cite{lyou1} appears to be a
suitable model to explain our data. In this approximation all the
spin components share the same spatial wavefunction
$\Phi(\textbf{r})$ and the total wavefunction is
$\Psi(\textbf{r},t)=\Phi(\textbf{r})(\sqrt{\rho_{-1}(t)}e^{i\theta_{-1}(t)},
\sqrt{\rho_{0}(t)}e^{i\theta_{0}(t)},\sqrt{\rho_{+1}(t)}e^{i\theta_{+1}(t)})$,
where $\theta_{m_F}$ represents the phase of each spin component.
Taking into account the conservation of $m$ and $N$, the
description simplifies into a model with only two dynamical
variables $\rho_0$ and $\theta$, where $\theta =
\theta_{-1}+\theta_{+1} - 2\theta_0$. The classical spinor energy
is~\cite{lyou1}
\begin{equation}
E=E_{\rm
qz}(1-\rho_0)+c\rho_0\left((1-\rho_0)+\sqrt{(1-\rho_0)^2-m^2}\cos\theta\right),\label{classicalE}
\end{equation}
where $E_{\rm qz}\propto B^2$ is the quadratic Zeeman shift
($E_{\rm qz}/h =(0.0277 \,\textrm{Hz}/( \mu\textrm{T})^2)B^2$),
$c= c_2 \langle n\rangle$ is the spin-dependent interaction energy
with the mean BEC density $\langle n\rangle$, and $c_2/h=
2.4\times10^{-13}$ Hz cm$^3$ for $^{23}$Na ($h$ is the Planck
constant)~\cite{lett}.  In the Thomas-Fermi approximation,
$\langle n\rangle$ and $c$ are proportional to
$N^{2/5}\omega_0^{6/5}$. The separatrix is the contour in
($\rho_0$, $\theta$) phase space with energy $E_{\rm sep}$, on
which there is a saddle point where $\dot{\rho_0}=\dot{\theta}=0$.
For an antiferromagnetic spinor BEC with $m=0$, $E_{\rm
sep}=E_{\rm qz}$. The mean-field ground state is $\rho_0=1$ for
$m=0$~\cite{faraday}.

The energy of the system cannot be directly inferred from SG-AI
measurements. We can, however, use the values of $\rho_0$ and $m$
to calculate an upper bound ($E_{\rm limit}$) to the classical
spinor energy. In fact, $E_{\rm limit}$ is equal to
$(1-\rho_{0}^{\rm max})(E_{\rm qz}+2c\rho_{0}^{\rm max})$ for
$m=0$, where $\rho_{0}^{\rm max}$ is the maximum value of $\rho_0$
among the 25-30 repeated measurements. The inset of
Fig.~\ref{qfactor} shows that $E_{\rm limit}$ decreases over time
and crosses the separatrix around $t_c$ at low fields. This
indicates that $E$ is not conserved.

One source of dissipation is loss of BEC atoms and the
corresponding decrease in density, resulting in a decrease in $c$.
The interaction energy evolves as $c(t)\propto e^{-2\gamma t/5}$
as follows from the Thomas-Fermi approximation, with $\gamma \leq$
0.02 s$^{-1}$ estimated from the observed atom loss. Because all
observed $t_c$ are much smaller than $1/\gamma$, the evolution of
$c$ does not explain our data and, hence, is ignored.

\begin{figure}[t]
\includegraphics[width=85mm]{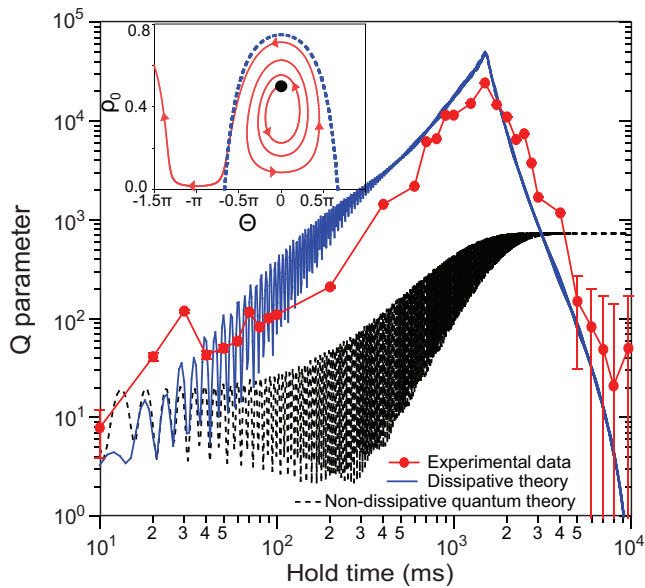}
\caption{(Color online) Time evolution of $Q$ with $\langle
n\rangle=2.02(8)\times10^{14}$ cm$^{-3}$ ($c/h =$ 48(2) Hz) at
$B=21.2(1)~\mu$T. Red dots represent experimental data. The dashed
(black) line is the result of the non-dissipative quantum
simulation. The solid (blue) line represents a classical Monte
Carlo simulation using Eqs.~(3) with $\beta\neq 0$. Inset: A
classical simulated path including energy dissipation through
phase space. The black dot represents the initial state. The
(blue) dashed line is the contour with energy $E_{\rm sep}$.}
\label{212mG}
\end{figure}

In Fig.~\ref{212mG} the evolution of $Q$ is compared to several
theoretical models. The dashed line is a result of a
non-dissipative quantum simulation based on a quantized version of
Eq.~\ref{classicalE} following the prescription in
Ref.~\onlinecite{youQ}. An initial Gaussian wavepacket with a
standard deviation in $\rho_0$ of 0.8\% mimics the fluctuations in
initial population. A classical Monte Carlo simulation based on
Eq.~\ref{classicalE} provides a similar result (not shown). We
average over 30 trajectories using a Gaussian probability
distribution for the same variance of $\rho_0$. These two
non-dissipative simulations approach an identical steady-state
value of $Q$, however, they do not explain the observed Q.
Moreover, the steady-state spin populations of these models are
different from experimental observations.

We modify the equations of motion for $\rho_0$ and
$\theta$~\cite{lyou1} by adding a dissipation term inspired by the
description of ohmic loss in Josephson junctions~\cite{kohler03}.
This leads to
\begin{eqnarray}
\dot{\rho_0}&=&
 -(2/\hbar)\partial E /\partial \theta,\nonumber\\
 \dot{\theta}&=& +(2/\hbar)  \partial E /\partial \rho_0+
 \beta\dot{\rho_0}\label{dissipation},
\end{eqnarray} which corresponds to an evolution of the energy
given by $dE/dt=-\hbar\beta(\dot{\rho_0})^2/2$. The solid line in
Fig.~\ref{212mG} represents a classical Monte Carlo simulation
with this dissipation mechanism. We sample two Gaussian
distributions with standard deviations of 0.8\% and 1\% for the
initial values of $\rho_0$ and $c$, respectively. The deviation in
$c$ is due to the 2\% uncertainty in $N$, while the drift in
$\omega_0$ over these measurements is negligible.

By assuming $E(t_c)=E_{\rm sep}$ the coefficient $\beta$ is
obtained, where $t_c$ is the observed critical time. For an
extracted $\beta$, we find agreement between the observed $E_{\rm
limit}$ and the energy derived from the dissipative model, as
shown in the inset of Fig.~\ref{qfactor}. Additionally, including
the initial fluctuations enables our model to reproduce the
behavior of $Q(t)$ in Fig.~\ref{212mG}. The predicted $Q$,
however, is almost twice as large as the experimental value at
$t_c$. Several other phenomenological dissipation terms were
tested, but the term $\beta\dot{\rho_0}$ is the only one linear in
$\rho_0$, $\theta$, $\dot{\rho_0}$, or $\dot{\theta}$ that we
found to drive the system to the correct ground
state~\cite{faraday}.

\begin{figure}[t]
\includegraphics[width=85mm]{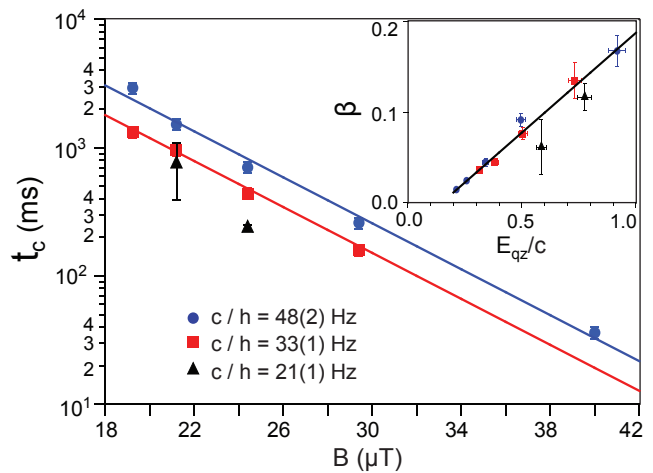}
\caption{(Color online) Measured $t_c$ as a function of $B$ at
mean densities, 2.02(8)$\times10^{14}$~cm$^{-3}$ (blue circles),
1.37(6)$\times10^{14}$~cm$^{-3}$ (red squares) and
0.89(4)$\times10^{14}$~cm$^{-3}$ (black triangles). The solid
lines are exponential fits. Inset: $\beta$ as a function of
$E_{\rm qz}/c$ for the same data. The solid line is a weighted
linear fit to the data. At the separatrix $E_{\rm qz}/c=1$.}
\label{beta}
\end{figure}

We can understand the evolution of $Q$ from considering the
trajectory in the inset of Fig.~\ref{212mG}. It shows a simulated
path through phase space based on Eqs.~\ref{dissipation}. The
dissipation leads to a gradual decrease of energy, which results
in a larger oscillation amplitude in $\rho_0$ as the trajectory
approaches the separatrix. Paths starting from slightly different
initial conditions separate after several oscillations. Larger
oscillation amplitudes in $\rho_0$ in these separated paths lead
to a larger variance and thus an increasing $Q$. As the energy
decreases past the separatrix, the oscillation amplitude decreases
until it goes to zero at the ground state. This reduces $Q$ back
to a minimum. Thus $Q$ reaches a maximum as the ensemble crosses
the separatrix at time $t_c$. The energy dissipation is
deterministic and does not explain the observed $Q$ by itself. The
evolution of $Q$ is a consequence of dephasing due to the spread
of initial conditions and the change in oscillation amplitude
during the evolution of the system.

Figure~\ref{beta} shows the measured $t_c$ as a function of $B$
for three densities (and thus for three values of $c$). The solid
lines show exponential fits to the data. Uncertainties are larger
for data at the smallest density because we find that $Q(t)$ shows
a broader and asymmetric peak and thus it is hard to extract $t_c$
from a single Gaussian fit to $Q(t)$.

The inset of Fig.~\ref{beta} shows $\beta$ determined from
Eqs.~\ref{dissipation} as a function of $E_{\rm qz}/c$. The
coefficient $\beta$ fits to a linear function with a
(dimensionless) slope of 0.22(1). The relation holds for different
$B$ and $\langle n\rangle$, hence, a single function seems to
describe all the data in terms of a single parameter $E_{\rm
qz}/c$. It is important to note that this same ratio completely
determines the shape of the energy surface in phase space for a
given $m$. For very small fields, the fit extrapolates to an
unphysical negative $\beta$. This indicates that either the
dissipation term or the functional form of $\beta$ are perhaps
incomplete or not appropriate. For example, our analysis does not
include the coupling of the single BEC mode to other degrees of
freedom, such as elementary excitations (whose energies are on the
order of $\hbar\omega_0$), which are indicated to be important by
Ref.~\onlinecite{MorenoCardoner}.

\begin{figure}[t]
\includegraphics[width=85mm]{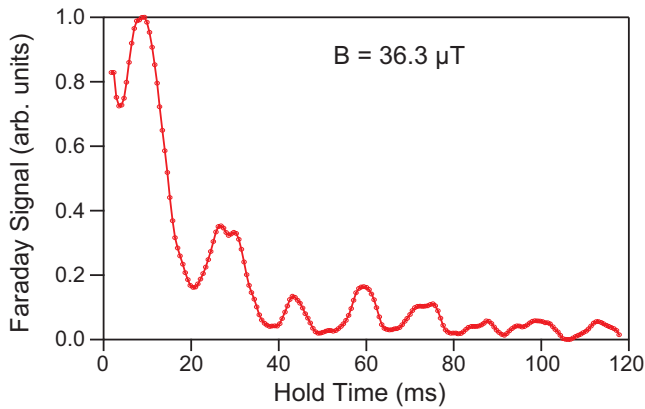}
\caption{Faraday signal for a single evolving BEC with $\langle
n\rangle=1.37(6)\times10^{14}$ cm$^{-3}$ ($c/h=$ 33(1) Hz). For this specific case, the initial $\rho_0$ is 0.45. The transition from the oscillating phase solution to the running phase solution happens between 35 ms and 45 ms, as shown by the signal periodically approaching zero at the minima afterwards.
This is in good agreement with the predicted $t_c$ from
Fig.~\ref{beta}.} \label{faraday}
\end{figure}

We have also studied the phase-space dynamics using Faraday
rotation spectroscopy. This method can reveal evidence of the
separatrix crossing from a single BEC realization. As outlined in
Ref.~\onlinecite{faraday}, a BEC on the high energy side of the
separatrix produces an oscillating Faraday signal with non-zero
minima, while on the low energy side the Faraday signal reaches
zero. Figure~\ref{faraday} shows an example of the Faraday signal
from a single BEC at $B=36.3(1)$~$\mu$T and $m=0$. The trace shows
a transition from non-zero minima to minima near zero, thus
providing a clear signature of crossing the separatrix in phase
space between 35~ms and 45~ms. While the details of repeated
traces vary, the system always crosses the separatrix during this
narrow time interval.

Figure~\ref{faraday} confirms our assumption that $t_c$
corresponds to a crossing of the separatrix, and $t_c$ from the
Faraday signal agrees with the extrapolated value of $t_c$ from
Fig.~\ref{beta}.  Light-induced atom loss limits our Faraday
detection to 100 ms, making it hard to observe the separatrix
crossing using Faraday detection for smaller values of $B$.

In Fig.~\ref{faraday} the rapid reduction of the oscillation
amplitude with time is due to atom losses generated by the Faraday
beam and not, as might be expected, to energy dissipation. The
Faraday beam gradually destroys the spin dynamics by off-resonant
light scattering and tensor light-shift dephasing, and thus the
Faraday detection is only effective over short observation times.

In conclusion, we have studied spin population fluctuations and
energy dissipation in a spinor BEC. Population fluctuations peak
at a critical time where the energy of the system equals that of
the separatrix in phase space and we have confirmed the separatrix
crossing using Faraday rotation spectroscopy. We present a
dissipation model with a single phenomenological coefficient that
describes our data. While the underlying physics requires further
study, this work sheds new light on dissipation mechanisms in a
nominally-isolated spinor BEC.

We thank W. D. Phillips, V. Boyer, and T. Hanna for insightful
discussions, and the ONR for financial support. SEM thanks the
NIST/NRC postdoctoral program. EG acknowledges support from
AMC-FUMEC.

\end{document}